\documentclass[conference]{IEEEtran}

\IEEEoverridecommandlockouts

\usepackage{xspace}
\usepackage{algorithm}
\usepackage{algpseudocode}
\usepackage{enumitem}
\usepackage{graphicx}
\usepackage{subcaption}
\usepackage{multirow}
\usepackage{booktabs}
\usepackage{array}
\usepackage{url}
\usepackage{xcolor}
\usepackage{amsmath}
\usepackage{amsfonts}
\usepackage{amssymb}
\usepackage{amsthm}

\usepackage{pifont}

\begin{document}

\newcommand{\red}[1]{\textcolor{red}{#1}}
\newcommand{\blue}[1]{\textcolor{blue}{#1}}
\newcommand{\purple}[1]{\textcolor{purple}{#1}}

\newcommand{\mostafa}[1]{\red{Mostafa:#1}}
\newcommand{\vlad}[1]{\blue{Vlad:#1}}
\newcommand{\todo}[1]{\purple{TODO:#1}}

\newcommand{\tnorm}{\theta}
\newcommand{\tconorm}{\kappa}
\newcommand{\fneg}{\eta}

\newcommand{\alc}{$\mathcal{ALC}$\xspace}

\newcommand{\baseEmb}{\texttt{baseline}\xspace}

\newcommand{\fProduct}{\texttt{Product}\xspace}
\newcommand{\fMinMax}{\texttt{Gödel}\xspace}
\newcommand{\fLuk}{\texttt{Łukasiewicz}\xspace}
\newcommand{\eSize}{\texttt{emb-size}\xspace}
\newcommand{\dProd}{\texttt{dot-product}\xspace}
\newcommand{\cosineS}{\texttt{cosine}\xspace}
\newcommand{\eucl}{\texttt{euclidean-distance}\xspace}

\newcommand{\vabbtwo}{ablation}
\newlength{\verticalsizeabbtwo}
\settowidth{\verticalsizeabbtwo}{\vabbtwo}
\newcommand{\abbtwostrut}{\rule[-0.25\verticalsizeabbtwo]{0pt}{0.75\verticalsizeabbtwo}}

\newcommand{\vintersection}{intersection}
\newlength{\verticalsizeintersection}
\settowidth{\verticalsizeintersection}{\vintersection}
\newcommand{\intersectionstrut}{\rule[-0.16\verticalsizeintersection]{0pt}{0.36\verticalsizeintersection}}

\newcommand{\uniontT}{\texttt{union}\xspace}
\newcommand{\ablationT}{\texttt{ablation}\xspace}
\newcommand{\subT}{\texttt{subsumption}\xspace}
\newcommand{\intersectT}{\texttt{intersection}\xspace}
\newcommand{\randW}{\texttt{random-walk}\xspace}

\newcommand{\plantO}{\texttt{PO}\xspace}
\newcommand{\idO}{\texttt{IDO}\xspace}

\newcommand{\mrr}{\texttt{MRR}\xspace}
\newcommand{\hitK}{\texttt{Hit@\(k\)}\xspace}
\newcommand{\overlapK}{\texttt{Overlap@\(k\)}\xspace}

\newcommand{\boxtheorem}{\hfill $\blacksquare$\vspace{2mm}}
\newcommand{\indep}{\perp}

\newcommand{\reg}{\mathcal{R}}
\newcommand{\regC}{\lambda}
\newcommand{\loss}{\mathcal{L}}
\newcommand{\lr}{\texttt{LR}\xspace}
\newcommand{\mlp}{\texttt{MLP}\xspace}
\newcommand{\hl}[1]{\textbf{#1}}      
\newcommand{\shl}[1]{\underline{#1}}  

\newcommand{\myparagraph}[1]{\vspace{2mm}\noindent \textit{#1.}}

\newcommand{\priCI}{\texttt{PrivCI}\xspace}
\newcommand{\privci}{\priCI}
\newcommand{\priCIG}{\priCI}
\newcommand{\otclean}{\texttt{OTClean}\xspace}
\newcommand{\ppgm}{\texttt{PrivatePGM}\xspace}
\newcommand{\prefair}{\texttt{PreFair}\xspace}

\newcommand{\kamino}{\texttt{Kamino}\xspace}
\newcommand{\original}
{\texttt{Original}\xspace}
\newcommand{\clean}
{\texttt{Clean}\xspace}
\newcommand{\dirty}
{\texttt{Dirty}\xspace}
\newcommand{\capuchin}{\texttt{Capuchin}\xspace}
\newcommand{\mst}{\texttt{MST}\xspace}
\newcommand{\dist}{\textit{dist}\xspace}

\newcommand{\car}{\texttt{Car}\xspace}
\newcommand{\adult}{\texttt{adult}\xspace}
\newcommand{\law}{\texttt{law}\xspace}
\newcommand{\boston}{\texttt{boston}\xspace}
\newcommand{\compas}{\texttt{compas}\xspace}
\newcommand{\dutch}{\texttt{dutch}\xspace}
\newcommand{\german}{\texttt{german}\xspace}

\newcommand{\X}{\red{X}\xspace}
\newcommand{\ten}[1]{\cdot 10^{#1}}
\newcommand{\pearson}{\texttt{Chi-Squared}\xspace}
\newcommand{\cmi}{\texttt{CMI}\xspace}
\newcommand{\MI}{\texttt{MI}\xspace}
\newcommand{\wdistance}{\texttt{W-Distance}\xspace}
\newcommand{\tvd}{\texttt{TV}\xspace}
\newcommand{\auc}{\texttt{AUC}\xspace}
\newcommand{\eqo}{\texttt{EO}\xspace}
\newcommand{\dpmetric}{\texttt{DP}\xspace}

\newcommand{\V}{\mathcal{V}}   
\newcommand{\E}{\mathcal{E}}   
\newcommand{\A}{\mathcal{A}}   
\newcommand{\Sset}{\mathcal{S}} 

\newcommand{\Sens}{S}          
\newcommand{\Y}{Y}             

\newcommand{\Mi}{M_i}          
\newcommand{\Mij}{M_{i,j}}     
\newcommand{\Mijbar}{\bar{M}_{i,j}} 
\newcommand{\Minoisy}{\tilde{M}_i}  
\newcommand{\Mijnoisy}{\tilde{M}_{i,j}} 

\newcommand{\qij}{q_{i,j}}     

\newcommand{\DSU}{\texttt{DSU}}                  
\newcommand{\EXPM}{\texttt{EXP-MECHANISM}}       
\newcommand{\NOPATH}{\texttt{NO-PROHIBITED-PATH}}

\newcommand{\set}[1]{\textbf{#1}}

\newcommand{\algCommentD}[1]{\hfill\texttt{\textcolor{blue}{\(\triangledown\)} \textcolor{gray}{\small #1}}}
\newcommand{\algCommentL}[1]{\hfill\texttt{\textcolor{blue}{\(\triangleleft\)} \textcolor{gray}{\small #1}}}

\captionsetup[subfigure]{font=small, skip=2pt}
\captionsetup[figure]{font=small}

\newcommand{\plotdir}{plots_boxplots}

\newcommand{\RowFour}[1]{%
\subcaptionbox{#1--KL\label{fig:#1-kl}}[0.24\linewidth]{\includegraphics[width=\linewidth]{\plotdir/#1_kl_boxplot_by_epsilon.pdf}}%
\hfill
\subcaptionbox{#1--TV\label{fig:#1-tv}}[0.24\linewidth]{\includegraphics[width=\linewidth]{\plotdir/#1_tv_boxplot_by_epsilon.pdf}}%
\hfill
\subcaptionbox{#1--q\label{fig:#1-q}}[0.24\linewidth]{\includegraphics[width=\linewidth]{\plotdir/#1_proxy_mi_sum_boxplot_by_epsilon.pdf}}%
\hfill
\subcaptionbox{#1--AUC\label{fig:#1-auc}}[0.24\linewidth]{\includegraphics[width=\linewidth]{\plotdir/#1_auc_boxplot_by_epsilon.pdf}}%
\\[-0.2em]
}

\newcommand{\RowTwo}[1]{%
\subcaptionbox{#1--CMI\label{fig:#1-cmi}}[0.49\linewidth]{\includegraphics[width=\linewidth]{\plotdir/#1_cmi_boxplot_by_epsilon.pdf}}%
\hfill
\subcaptionbox{#1--EO\label{fig:#1-eo}}[0.49\linewidth]{\includegraphics[width=\linewidth]{\plotdir/#1_eo_boxplot_by_epsilon.pdf}}%
\\[-0.2em]
}

\title{Structure Selection for Fairness-Constrained Differentially Private Data Synthesis}

\author{%
\IEEEauthorblockN{Naeim Ghahramanpour}
\IEEEauthorblockA{University of Western Ontario\\
London, Canada\\
nghahram@uwo.ca}
\and
\IEEEauthorblockN{Mostafa Milani}
\IEEEauthorblockA{University of Western Ontario\\
London, Canada\\
mostafa.milani@uwo.ca}
}

\maketitle

\begin{abstract}
\emph{Differential privacy} (DP) enables safe data release, with synthetic data generation emerging as a common approach in recent years. Yet standard synthesizers preserve all dependencies in the data, including spurious correlations between sensitive attributes and outcomes. In fairness-critical settings, this reproduces unwanted bias. A principled remedy is to enforce \emph{conditional independence} (CI) constraints, which encode domain knowledge or legal requirements that outcomes be independent of sensitive attributes once admissible factors are accounted for.

DP synthesis typically proceeds in two phases: (i) a measurement step that privatizes selected marginals, often structured via maximum spanning trees (MSTs), and (ii) a reconstruction step that fits a probabilistic model consistent with the noisy marginals. We propose \priCI, which enforces CI during the measurement step via a CI-aware greedy MST algorithm that integrates feasibility checks into Kruskal’s construction under the exponential mechanism, improving accuracy over competing methods.

Experiments on standard fairness benchmarks show that \priCI achieves stronger fidelity and predictive accuracy than prior baselines while satisfying the specified CI constraints.
\end{abstract}


\theoremstyle{definition}
\newtheorem{defn}{Definition}
\theoremstyle{plain}
\newtheorem{thm}{Theorem}

\section{Introduction} \label{sec:intro}

\textit{Differential Privacy (DP)}~\cite{dwork2006differential, dwork2008differential} has emerged as the de facto standard for rigorous privacy guarantees in data analysis and machine learning. A prominent instantiation is \emph{DP synthetic data generation}~\cite{hu2024sok, fan2020survey, mckenna2019graphical, mckenna2021winning}, where one learns a noisy approximation of the data distribution under DP and samples artificial records from it, enabling broad downstream analysis from a single privacy-protected artifact.

While DP synthetic data enables broad downstream analysis, it is designed to preserve the statistical structure of the original dataset—up to controlled noise. Consequently, synthetic data also inherits the biases present in the source data. Real-world datasets often encode historical inequities or sampling artifacts, which can give rise to \emph{spurious correlations}—statistical dependencies that do not reflect true causal relationships in the domain~\cite{ye2024spurious, izmailov2022feature, salimi2019capuchin, pirhadi2024otclean}. In fairness-critical settings, these correlations may link protected attributes to outcomes in undesirable ways. For example, in hiring data, applicants from certain groups may have historically received fewer callbacks despite comparable qualifications~\cite{bertrand2004emily}, creating an association between sensitive attributes and hiring outcomes that does not reflect admissible decision criteria. When DP synthesizers preserve such patterns, they can replicate unfair dependencies in the released synthetic data, undermining both fairness and the validity of downstream analyses.

Importantly, DP provides guarantees about individual privacy—not about which dependencies should or should not be preserved. Consequently, a synthesizer that faithfully captures the data distribution may also preserve spurious or policy-prohibited correlations. A principled way to control which dependencies are retained is to encode domain or legal requirements as \emph{conditional independence (CI) constraints}. A CI constraint $X \indep Y \mid Z$ formalizes the requirement that outcomes $Y$ be independent of sensitive attributes $X$ once admissible factors $Z$ are accounted for. In the hiring example, this can be expressed as
$\texttt{Gender} \!\indep\! \texttt{Outcome} \mid \texttt{Qualifications}$,
ensuring that decisions are driven by qualifications rather than gender.

In this work, we study the problem of \emph{differentially private data synthesis under conditional independence (CI) constraints}. Our goal is to enforce CI requirements \emph{within} the DP synthesis pipeline, rather than applying post-hoc repair to already synthesized data. Integrating structural constraints directly into the private measurement process is crucial: DP mechanisms allocate privacy budget to estimate and preserve dependencies, and if some dependencies are policy-prohibited, preserving them and later correcting them can waste privacy budget and amplify distortion.

Most DP synthesizers for structured tabular data follow a two-stage \emph{measurement--reconstruction} paradigm~\cite{hu2024sok, mckenna2021winning, mckenna2019graphical}. In the \textit{measurement} stage, the algorithm privately estimates a collection of low-dimensional statistics (e.g., marginals or dependency scores). In the \textit{reconstruction} stage, it fits a joint distribution consistent with these noisy measurements and samples synthetic records. This modular decomposition enables efficient privacy accounting and scalability to moderate dimensions.

Within this framework, \mst~\cite{mckenna2021winning} and \ppgm~\cite{mckenna2019graphical} form a widely used backbone. \mst selects a dependency tree by privately ranking candidate edges using the exponential mechanism, and \ppgm reconstructs a graphical model from the selected marginals. However, standard MST-based pipelines optimize purely for dependency strength and do not account for structural constraints such as CI requirements. As a result, they may allocate privacy budget to dependencies that violate policy or fairness specifications.

To formalize such structural constraints, we adopt a fairness-oriented partition of attributes that is standard in causal and fairness-aware modeling~\cite{nabi2018fair, zhang2018mitigating, salimi2019capuchin}. 
Let $V$ denote the full set of attributes in the dataset. We divide $V$ into four roles: 
\emph{protected} attributes $S$ (e.g., gender or race), 
\emph{outcome} attributes $O$ (e.g., hiring decision or credit risk), 
\emph{admissible} attributes $A$ (variables through which dependence between $S$ and $O$ is considered legitimate), 
and optionally \emph{inadmissible} attributes $I$ (remaining covariates), 
so that $V = S \cup O \cup A \cup I$.

The distinction between admissible and inadmissible attributes captures a common policy intuition: some pathways from $S$ to $O$ may be acceptable because they operate through relevant qualifications (e.g., education, experience, or test scores), whereas other pathways reflect direct or proxy discrimination and should be eliminated. 
For instance, in hiring, $\texttt{Qualifications}$ may be treated as admissible, while variables that act as proxies for protected status—such as $\texttt{ZipCode}$, $\texttt{Neighborhood}$, or prior $\texttt{Employer}$ in settings where it encodes segregation effects—may be considered inadmissible mediators. 
Admissible variables therefore represent factors through which dependence is permitted, while inadmissible variables represent channels through which protected-to-outcome influence is not allowed.

A fairness requirement is then expressed as a conditional independence (CI) constraint of the form
\[
O \indep S \mid A,
\]
which requires that once admissible factors are fixed, the outcome be statistically independent of the protected attributes.

In graphical models, CI constraints correspond to separation properties~\cite{koller2009probabilistic}: 
$O \indep S \mid A$ holds if $A$ separates $S$ and $O$ in the graph. 
In a tree-structured dependency model, this reduces to a structural requirement: 
any path connecting a node in $S$ to a node in $O$ must pass through at least one node in $A$. 
Equivalently, in the graph induced by $V \setminus A$, there must be no path between $S$ and $O$.

We propose \priCI, a DP synthesis framework that enforces CI constraints 
\emph{structurally during the measurement stage}. 
\priCI\ modifies the Kruskal-style maximum spanning tree construction in \mst 
by treating the CI requirement as a \emph{feasibility constraint} during edge selection. 
At each iteration, a candidate edge is considered admissible only if 
(i) it preserves acyclicity and 
(ii) its addition does not introduce an $S$--$O$ path in the graph induced by $V \setminus A$. 
Equivalently, we require that $A$ separates $S$ and $O$ in the learned tree. 
Edges that would violate this separation condition are excluded. 
Among the remaining feasible edges, selection proceeds privately via the exponential mechanism.

The resulting algorithm is a greedy, CI-aware marginal selection procedure that retains the privacy accounting and composition properties of the underlying MST pipeline, while ensuring that the learned dependency structure satisfies the specified fairness constraint.

Our approach emphasizes practical utility under DP budgets while structurally enforcing CI during marginal selection. We focus on empirical performance and fairness outcomes across benchmarks rather than additional theoretical guarantees.

We evaluate \priCI\ on standard fairness benchmarks across multiple privacy budgets. We compare against an existing CI-constrained DP synthesis method~\cite{pujol2022prefair}. Across datasets, \priCI\ improves distributional fidelity and downstream predictive performance while satisfying the specified CI constraints, demonstrating that enforcing structural separation during measurement yields a favorable privacy--utility trade-off.

The remainder of the paper is organized as follows. Section~\ref{sec:relatedwork} discusses related work, Section~\ref{sec:background} reviews relevant background, Section~\ref{sec:methods} presents the proposed methods, Section~\ref{sec:experiments} reports experimental results, and Section~\ref{sec:conclusion} concludes.

\section{Related Works}
\label{sec:relatedwork}

To position our contributions, we first note that existing non-private CI repair methods, such as \capuchin~\cite{salimi2019capuchin} and \otclean~\cite{pirhadi2024otclean}, modify the data distribution to satisfy fairness constraints by altering tuples or reweighting probability mass. However, combining such post-hoc repair with DP synthesis is suboptimal. Repairing after synthesis ignores distortions already introduced by DP, while repairing before synthesis may waste privacy budget preserving correlations that will later be removed. In contrast, \priCI integrates fairness directly into the DP synthesis process by constraining dependency selection from the outset.

Beyond synthetic data generation, several works have explored the intersection of privacy, data quality, and fairness in data management systems. For example, PACAS and its extension \cite{huang2018pacas,huang2020privacy} study privacy-aware data cleaning services, where data quality operations must be performed under privacy constraints. While these systems focus on cleaning inconsistent or noisy data rather than enforcing statistical independence constraints, they highlight the broader challenge of integrating data quality objectives with privacy guarantees. Related concerns also arise in fairness-aware data integration tasks such as entity matching, where bias in matching decisions can propagate to downstream analyses. For instance, recent work on threshold-independent fair matching \cite{moslemi2024threshold,nilforoushan2022entity} proposes score calibration techniques to mitigate unfairness in entity matching pipelines. These works illustrate how privacy, fairness, and data quality considerations increasingly intersect across different stages of the data lifecycle.

This section reviews related work on DP synthetic data generation and on enforcing semantic or statistical constraints under DP.

\subsection{Differentially Private Synthetic Data Generation}
\label{sec:rw-dynthetic}

Differentially private (DP) synthetic data generation produces a new dataset that preserves statistical properties of the original data under formal privacy guarantees. Hu et al.~\cite{hu2024sok} note a common two-step paradigm: privatize a compact summary (e.g., marginals or dependencies) and reconstruct a joint distribution to sample synthetic records.

For tabular data, many methods rely on low-dimensional marginals. \emph{PrivBayes}~\cite{zhang2017privbayes} learns a Bayesian network from noisy marginals. Graphical-model approaches such as \ppgm~\cite{mckenna2019graphical}, \emph{PrivMRF}~\cite{cai2021data}, and \emph{AIM}~\cite{mckenna2022aim} select informative marginals and fit consistent models, with AIM adapting selection via the exponential mechanism. \emph{JAM-PGM}~\cite{fuentes2024joint} incorporates public data to reduce private measurements.

Other settings include local DP (e.g., \emph{PrivIncr}~\cite{liu2023multi}) and DP deep generative models via DP-SGD~\cite{abadi2016deep} (e.g., DPGAN, DP-CGAN, PATE-GAN), though these often face stability and utility challenges. Dataset condensation methods (e.g., Dyn-PSG~\cite{wu2024dynamic}) provide another direction but primarily optimize predictive accuracy rather than enforcing structural constraints. Overall, prior DP synthesizers focus on fidelity under privacy; our work adds explicit CI constraints during synthesis.

\subsection{Enforcing Constraints with DP Synthetic Data}
\label{sec:rw-constraint}

Real datasets may be incomplete or violate structural constraints. Mohapatra et al.~\cite{mohapatra2023differentially} integrate imputation with DP synthesis by adapting PrivBayes and MisGAN~\cite{li2019misgan}. \kamino~\cite{ge2020kamino} enforces integrity constraints (e.g., denial constraints, functional dependencies) by learning a private Bayesian network and sampling with a constraint-aware Markov chain. These approaches target logical constraints rather than statistical constraints such as CI.

\subsection{\prefair: Enforcing CI via Structural Repair}
\label{sec:prefair}

\prefair~\cite{pujol2022prefair} is the closest prior work to ours. It targets \emph{justifiable fairness}, which is motivated by causal reasoning. The key requirement is that any statistical influence of protected attributes on an outcome must be mediated through admissible attributes. Equivalently, in an appropriate causal or dependency graph, every directed path from a protected attribute to an outcome must pass through at least one admissible attribute. This blocks direct or inadmissible routes from protected to outcome and aims to ensure that any remaining dependence is explainable through acceptable mediators.

Operationally, \prefair enforces this requirement by constraining the dependency structure used for marginal-based synthesis. It constructs a fair Marginals-MST, a maximum spanning tree over attributes where edge weights reflect pairwise dependence. The fairness constraint restricts which spanning trees are allowed: the chosen tree must satisfy the admissibility separation condition between protected and outcome attributes. This structural restriction is then used to justify that the induced distribution satisfies justifiable fairness under the intended interpretation of the tree as a dependency graph.

To obtain such a fair tree, \prefair uses a scalable heuristic that modifies a private Kruskal-style procedure by removing or avoiding inadmissible connections during construction (for example, edges that would create an inadmissible protected--outcome path). This improves scalability but may sacrifice utility because the resulting tree can be suboptimal among fair trees. An exhaustive variant instead searches for the best fair tree using a Gaussian-mechanism-based Kruskal procedure; it tends to underperform \prefair and \priCI in practice and is therefore excluded from our experiments.

After a fair dependency tree is obtained, \prefair proceeds with private parameter estimation and sampling. It privately measures the required 1-way marginals and the 2-way marginals corresponding to the tree edges, and then uses \ppgm~\cite{mckenna2019graphical} to fit a graphical model consistent with these privatized measurements. Synthetic data is produced by sampling from the fitted model. \prefair provides formal guarantees for both fairness and privacy: it shows that if the learned Marginals-MST satisfies the stated topological condition, then the resulting distribution is justifiably fair, and it proves that the overall procedure satisfies Rényi DP (and thus can be converted to $(\epsilon,\delta)$-DP) under standard composition.
\section{Background} \label{sec:background}

We briefly review the key concepts that underpin our methods: DP, CI, \mst, and \ppgm.

\subsection{Foundations of Differential Privacy} \label{sec:DP}

\emph{Differential privacy} (DP)~\cite{dwork2006calibrating, dwork2014algorithmic} ensures that the output of a randomized algorithm is nearly indistinguishable on neighboring datasets differing in one individual. 
Formally, a mechanism $\mathcal{M}$ satisfies $(\epsilon,\delta)$-DP if, for all neighboring $D,D'$ and measurable sets $S$,
\begin{align}
P(\mathcal{M}(D)\in S) \leq e^\epsilon P(\mathcal{M}(D')\in S) + \delta.
\end{align}
Here $\epsilon$ controls the privacy loss, and $\delta$ bounds the probability of rare failure events. The special case $\delta=0$ is called \emph{pure} DP.

\emph{Rényi differential privacy (RDP)}~\cite{mironov2017renyi} refines DP in a way that is particularly convenient for composition. 
A mechanism $\mathcal{M}$ satisfies $(\alpha,\gamma)$-RDP if, for all neighboring $D,D'$,
\begin{align}
d_\alpha\!\left(\mathcal{M}(D)\,\|\,\mathcal{M}(D')\right) \leq \gamma,
\end{align}
where $d_\alpha$ denotes Rényi divergence of order $\alpha>1$. 
RDP composes additively: if mechanisms $\mathcal{M}_1,\dots,\mathcal{M}_k$ satisfy $(\alpha,\gamma_i)$-RDP, their composition satisfies $(\alpha,\sum_i \gamma_i)$-RDP.

A common specialization is \emph{zero-concentrated DP (zCDP)}~\cite{bun2016concentrated}, where a mechanism satisfies $\rho$-zCDP if it is $(\alpha,\alpha\rho)$-RDP for all $\alpha>1$. 
zCDP provides a single parameter $\rho$ that composes linearly and admits tight conversion to $(\epsilon,\delta)$-DP. 
Specifically, a $\rho$-zCDP mechanism satisfies $(\epsilon,\delta)$-DP for any $\delta>0$ with
\begin{align}
\epsilon = \rho + 2\sqrt{\rho \log(1/\delta)}.
\end{align}
Conversely, for a target $(\epsilon,\delta)$ guarantee, it is convenient to work with the equivalent zCDP budget
\begin{align}
\rho 
= \big(\sqrt{\log(1/\delta)+\epsilon}-\sqrt{\log(1/\delta)}\big)^2,
\label{fr:conversion}
\end{align}
which we use to calibrate the per-stage budgets in \mst\ and \priCI.

Differential privacy is achieved by adding calibrated noise based on a query’s sensitivity. 
For $f:\mathcal{D}\to\mathbb{R}^k$, the $\ell_p$-sensitivity is
\begin{align}
\Delta_p f = \max_{D\sim D'} \|f(D)-f(D')\|_p.
\end{align}

The primitives most relevant to this work are:

\begin{itemize}
\item \emph{Gaussian mechanism.} Adds Gaussian noise scaled to $\ell_2$-sensitivity. 
For sensitivity~1, adding $\mathcal{N}(0,\sigma^2)$ noise satisfies $(\alpha,\tfrac{\alpha}{2\sigma^2})$-RDP, equivalently $\rho$-zCDP with $\rho=\tfrac{1}{2\sigma^2}$. 
This mechanism is used for privately measuring marginals.

\item \emph{Exponential mechanism.} 
Given a quality score $q(D,r)$ with sensitivity $\Delta q$, the mechanism samples $r$ with probability proportional to
\begin{align}
\exp\!\left(\tfrac{\epsilon}{2\Delta q}\, q(D,r)\right).
\end{align}
It provides $\epsilon$-DP and is the standard tool for private selection, including edge selection in \mst.
\end{itemize}

\subsection{Conditional Independence} \label{sec:ci}

A conditional independence (CI) constraint $\sigma$ takes the form $\set{X} \perp \set{Y} \mid \set{Z}$, meaning that
\[
P(\set{X},\set{Y}\mid\set{Z}) 
= P(\set{X}\mid\set{Z})P(\set{Y}\mid\set{Z}).
\]
CI constraints are fundamental in statistics and graphical modeling, and in our setting they encode fairness requirements such as outcomes being independent of protected attributes given admissible factors.

From finite data, CI can be quantified using \emph{conditional mutual information} (CMI),
\[
I(\set{X};\set{Y}\mid\set{Z})
= \sum_{x,y,z} 
P(x,y,z)\,
\log \frac{P(x,y\mid z)}{P(x\mid z)P(y\mid z)}.
\]
CMI is nonnegative and equals zero if and only if $\sigma$ holds. We use CMI as a quantitative measure of CI violation in synthetic data.

\subsection{\ppgm and \mst} \label{sec:MST}

\ppgm~\cite{mckenna2019graphical} reconstructs a joint distribution from a collection of privatized low-dimensional marginals. Given noisy 1-way marginals and a selected set of 2-way marginals, \ppgm fits a graphical model consistent with these measurements and enables sampling of synthetic records. In this work, \ppgm is used as an off-the-shelf reconstruction module.

A common paradigm in DP data synthesis is to measure a restricted set of low-dimensional marginals and reconstruct an approximate joint distribution. Since measuring all marginals is infeasible in high dimensions, algorithms such as \mst~\cite{mckenna2021winning} privately select a compact, informative subset.

A widely used structural restriction is to approximate the joint distribution by a \emph{tree-structured graphical model}. Tree structures require only $d-1$ pairwise marginals, admit efficient inference, and provide a principled tradeoff between expressiveness and scalability. Selecting such a structure reduces to computing a maximum spanning tree over attributes with respect to a dependency score.

Let $\mathcal{T}_d$ denote the set of spanning trees over $d$ attributes. 
For an edge $e_{i,j}$ between attributes $(X_i,X_j)$, let $q(e_{i,j}) \ge 0$ denote a dependency score. 
Selecting a tree-structured approximation then reduces to solving
\begin{align}
T_q^* 
= \arg\max_{T \in \mathcal{T}_d} 
\sum_{e_{i,j}\in T} q(e_{i,j}),
\label{eq:weight-q}
\end{align}
that is, finding a maximum spanning tree under the edge weights $q$.

Different choices of $q$ induce different approximations and privacy--utility tradeoffs. 
A classical special case is the Chow--Liu algorithm~\cite{chow1968approximating}, which sets 
$q(e_{i,j}) = \MI(X_i;X_j)$, the mutual information between $X_i$ and $X_j$, defined as
\[
\MI(X_i;X_j)
= \sum_{x_i,x_j} 
P(x_i,x_j)\,
\log \frac{P(x_i,x_j)}{P(x_i)P(x_j)}.
\]
The resulting tree maximizes total mutual information and yields the tree-structured distribution
\[
P_T(x_1,\dots,x_d)
= \prod_{i=1}^d P(X_i \mid X_{\pi(i)})
\]
that minimizes $\mathrm{KL}(P\,\|\,P_T)$ among all tree-structured models.

However, mutual information is not well suited for private edge scoring because its global sensitivity can be large. 
Accordingly, \mst employs lower-sensitivity discrepancy scores, such as
\[
q(e_{i,j}) 
= \|P(X_i,X_j) - P(X_i)P(X_j)\|_2^2,
\]
which vanish under independence and increase with dependence strength.

The \mst algorithm instantiates a differentially private version of this tree-selection procedure. 
It first measures all $d$ one-way marginals using the Gaussian mechanism under a portion of the privacy budget. 
Next, for each candidate edge $e_{i,j}$, a dependency score $q_{i,j}$ is computed from noisy 2-way marginals. 
A maximum spanning tree is then constructed using the exponential mechanism,
\begin{align}
\Pr(e_{i,j}) 
\propto 
\exp\!\left(
    \tfrac{\varepsilon}{2\Delta q}\, q_{i,j}
\right),
\end{align}
which ensures differential privacy while favoring edges with stronger estimated dependence. 
Finally, each selected edge $e_{i,j} \in T$ is re-measured with the Gaussian mechanism, and the resulting noisy 1-way and 2-way marginals are passed to \ppgm\ for reconstruction and sampling.

The overall pipeline satisfies $(\varepsilon,\delta)$-DP under standard composition. 
Because only $O(d)$ marginals are measured, inference and reconstruction scale efficiently to moderate and high dimensions. 
In the non-private limit, the procedure reduces to a maximum spanning tree over dependency scores; under privacy noise, it provides a principled tree-structured approximation with formal privacy guarantees.

\section{DP Data Synthesis with CI Constraints}
\label{sec:methods}

This section introduces \priCI, our framework for generating DP synthetic data while enforcing user-specified CI constraints. CI constraints help eliminate spurious or unwanted correlations—especially between sensitive attributes and outcomes—so that the synthetic data is both privacy-preserving and fairness-aware. \priCI extends the standard two-stage DP synthesis pipeline (\mst for marginal selection and \ppgm for model fitting) by enforcing CI structurally during the marginal-selection stage in \mst.

\subsection{Problem Definition} \label{sec:ps}

Let $D$ be a dataset over attributes $\set{V} = \{X_1,\dots,X_d\}$, and let 
$\sigma : \set{X} \perp \set{Y} \mid \set{Z}$ be a user-specified conditional independence constraint with 
$\set{X},\set{Y},\set{Z} \subseteq \set{V}$. Our goal is to design a randomized mechanism $\mathcal{M}$ that produces a synthetic dataset 
$\tilde D = \mathcal{M}(D)$ satisfying the following requirements:

\begin{defn}[DP Data Synthesis with CI Constraints]
A mechanism $\mathcal{M}$ satisfies DP data synthesis with CI constraint $\sigma$ if:

\begin{enumerate}
    \item $\mathcal{M}$ satisfies $(\epsilon,\delta)$-differential privacy;
    \item the distribution $P_{\mathcal{M}(D)}$ induced by the learned model satisfies the CI constraint $\sigma$;
    \item the synthetic distribution remains close to the data distribution under a chosen utility metric.
\end{enumerate}
\end{defn}

Condition (ii) is enforced at the level of the learned graphical model. 
Condition (iii) is treated as an optimization objective rather than a hard constraint; in practice, utility is measured using statistical distances (e.g., KL divergence or statistical distances) and downstream predictive performance.

\subsection{\priCI: Greedy Structure Selection}
\label{sec:privCIG}

\priCI modifies the marginal-selection stage of \mst\ by enforcing a structural CI constraint during tree construction. 
The overall pipeline follows \mst, except that candidate edges are additionally filtered to ensure compliance with a user-specified constraint 
$\sigma: \set{X} \perp \set{Y} \mid \set{Z}$.

Given a target $(\varepsilon,\delta)$ guarantee, we first convert it to an equivalent $\rho$-zCDP budget using~\eqref{fr:conversion}. 
The total budget $\rho$ is divided evenly into three parts, as in \mst: 
(i) measuring all 1-way marginals, 
(ii) selecting tree edges via the exponential mechanism, and 
(iii) measuring the selected 2-way marginals. 
Each group of marginals is released in a single Gaussian mechanism call with appropriately normalized sensitivity so that each stage consumes $\rho/3$. 
For edge selection, the exponential mechanism is applied for $d-1$ iterations with per-iteration parameter $\epsilon'$, calibrated so that the total RDP cost equals $\rho/3$ under composition.

Let $E$ denote the current edge set. 
At each iteration, we consider candidate edges $(i,j)$ that preserve acyclicity. 
In addition, we exclude any edge whose addition would violate the CI constraint. 
Specifically, an edge is deemed infeasible if adding it would create a path between $\set{X}$ and $\set{Y}$ in the graph induced by $V \setminus \set{Z}$. 
Equivalently, we require that $\set{Z}$ separate $\set{X}$ and $\set{Y}$ in the evolving tree. 

Formally, the feasible candidate set at iteration $t$ is $S_t 
= \{ (i,j) : 
\texttt{IsTree}(E\cup (i,j)) \wedge \texttt{IsCIConsistent}(E\cup (i,j),\sigma_{\text{CI}}) \}$.
An edge is then sampled from $S_t$ using the exponential mechanism with score $q_{i,j}$. 
After $d-1$ edges are selected, the corresponding 2-way marginals are measured with the Gaussian mechanism and passed, together with the noisy 1-way marginals, to \ppgm\ for reconstruction and sampling.

Relative to \mst, the only modification is the feasibility filter 
\texttt{IsCIConsistent}. 
All privacy guarantees follow directly from those of \mst\ under the same composition analysis.

\begin{algorithm}[H]
\caption{\priCI}
\label{alg:pricig}
\begin{algorithmic}[1]
\State \textbf{Input:} Dataset $D$ over attributes $\{X_1,\dots,X_d\}$, CI constraint $\sigma_{\text{CI}}: \set{X}\perp\set{Y}\mid\set{Z}$, privacy $(\varepsilon,\delta)$
\State \textbf{Output:} Synthetic dataset $\tilde D$

\State $\rho \leftarrow \big(\sqrt{\log(1/\delta)+\varepsilon}-\sqrt{\log(1/\delta)}\big)^2$ \Comment{Set total zCDP budget} \label{ln:conversion}
\State $\rho' \leftarrow \rho/3$ \Comment{Per-stage zCDP budget}
\State $\sigma_G \leftarrow 1/\sqrt{2\rho'}$ \Comment{Gaussian scale (unit $\ell_2$ sensitivity)}
\State $\epsilon' \leftarrow \sqrt{8\rho'/(d-1)}$ \Comment{EM parameter}

\State $\{\tilde \mu_{\{i\}} : i \in [1,d]\} 
\leftarrow \texttt{OneWayMargs}(D,\sigma_G)$ \label{ln:oneway}

\State $\{q_{\{i,j\}} : i,j \in [1,d]\} 
\leftarrow \texttt{QualityScores}(D,\{\tilde \mu_{\{i\}}\})$

\State $E \leftarrow \emptyset$ \label{ln:beginMST}

\For{$t=1$ \textbf{to} $d-1$}
  \State $S \leftarrow \{(i,j): \texttt{IsTree}(E\cup (i,j))$
  \State \hspace{1cm} $\wedge \texttt{IsCIConsistent}(E\cup (i,j),\sigma_{\text{CI}})\}$
  \State $(u,v) \leftarrow \texttt{ExponentialMechanism}(S,q,\epsilon')$
  \State $E \leftarrow E \cup \{(u,v)\}$
\EndFor

\State $\{\tilde \mu_{\{i,j\}} : (i,j)\in E\} 
\leftarrow \texttt{TwoWayMargs}(D,E,\sigma_G)$ \label{ln:twoway}

\State $\hat P \leftarrow \ppgm(\tilde \mu)$
\State $\tilde D \leftarrow \texttt{Sample}(\hat P)$
\State \Return $\tilde D$
\end{algorithmic}
\end{algorithm}
\section{Experiments}\label{sec:experiments}

We pursue two main objectives in our experiments. First, we demonstrate the trade-off between data utility, privacy, and fairness. Utility is measured by the sum of $q$ (as a proxy for mutual information), statistical distribution fidelity, and downstream model performance (AUC). Privacy is controlled by the privacy budget $\epsilon$. Fairness is evaluated using conditional mutual information (CMI) and equalized odds (EO) in the downstream model. Second, we show that \privci\ achieves a better balance between utility and fairness compared to the competing method \prefair. In particular, it preserves higher utility while enforcing the fairness conditional independence constraint.

We first describe the experimental setup, including the datasets and data preparation. We then present the experimental results in Section~\ref{sec:results}, followed by analysis and key takeaways in Section~\ref{sec:discussion}.

\subsection{Experimental Setup}

We evaluate \priCI on five benchmark datasets widely used in algorithmic fairness research: 
Adult~\cite{dua2019uci}, 
COMPAS~\cite{angwin2016machine}, 
Dutch Census~\cite{kamiran2009classifying}, 
German Credit~\cite{dua2019uci}, 
and Law School~\cite{wightman1998lsac}. 
All datasets are used in their pre-encoded tabular form.

For each dataset, attributes are partitioned into outcome ($O$), 
protected ($S$), 
admissible ($A$), 
and inadmissible ($I$) variables, and we enforce the CI constraint 
$O \perp S \mid A$ during synthetic data generation.

\begin{itemize}[leftmargin=*]

\item \textbf{Adult.} 
Outcome: \texttt{income}. 
Protected: \texttt{gender}. 
Admissible: \texttt{occupation}, \texttt{education}, \texttt{hours-per-week}. 
Inadmissible: \texttt{marital-status}, \texttt{age}. 
The outcome indicates whether annual income exceeds \$50K.

\item \textbf{COMPAS.} 
Outcome: \texttt{is\_recid}. 
Protected: \texttt{race}. 
Admissible: \texttt{age}. 
Inadmissible: \texttt{charge\_degree}, \texttt{priors\_count}. 
The outcome indicates two-year recidivism.

\item \textbf{Dutch Census.} 
Outcome: \texttt{occupation}. 
Protected: \texttt{sex}. 
Admissible: \texttt{edu\_level}. 
Inadmissible: \texttt{economic\_status}, \texttt{cur\_eco\_activity}, 
\texttt{household}, \texttt{age}, \texttt{marital\_status}, 
\texttt{country\_birth}, \texttt{citizenship}, \texttt{prev\_residence}.

\item \textbf{German Credit.} 
Outcome: \texttt{risk}. 
Protected: \texttt{sex}. 
Admissible: \texttt{duration}, \texttt{credit\_amount}. 
Inadmissible: \texttt{age}, \texttt{job}, \texttt{housing},\\
\texttt{saving\_accounts}, \texttt{checking\_account}, \texttt{purpose}.

\item \textbf{Law School.} 
Outcome: \texttt{pass\_bar}. 
Protected: \texttt{race}. 
Admissible: \texttt{lsat}, \texttt{ugpa}. 
Inadmissible: \texttt{decile*}, \texttt{zfygpa}, \texttt{zgpa}, 
\texttt{fulltime}, \texttt{fam\_inc}, \texttt{male}, \texttt{tier}.

\end{itemize}

Continuous attributes are discretized prior to private synthesis using dataset-specific binning schemes described in Appendix~X. 
All evaluation metrics are computed on the discretized representation to ensure consistency between real and synthetic data.

\begin{table}[t]
\centering
\renewcommand{\arraystretch}{1.05}
\setlength{\tabcolsep}{2.0mm}
\begin{tabular}{lcrr}
\toprule
\textbf{Name} & \textbf{Key} & \textbf{\#Rows} & \textbf{\#Cols} \\
\midrule
Adult
& \adult
& 48{,}842
& 7
\\

COMPAS
& \compas
& 7{,}214
& 5
\\

Dutch Census
& \dutch
& 60{,}420
& 12
\\

German Credit
& \german
& 1{,}000
& 10
\\

Law School
& \law
& 20{,}798
& 12
\\
\bottomrule
\end{tabular}
\caption{Dataset properties used in our experiments. Row/column counts correspond to the encoded input tables; role assignments (outcome/protected/admissible/inadmissible) are described in the text.}
\label{tab:dataset-properties}
\end{table}

We compare with \prefair (Section~\ref{sec:prefair}) as the only existing method for enforcing similar constraints in data synthesis. We also include \mst (Section~\ref{sec:MST}) as a baseline which does not enforce fairness constraints but provide a baseline for comparing the performance.

\subsection{Experimental Results}
\label{sec:results}

We organize the evaluation into four parts. Section~\ref{sec:q} compares structural optimization performance via the proxy MI objective. Sections~\ref{sec:fidelity} and~\ref{sec:ml} evaluate distributional fidelity and downstream predictive utility, respectively. Section~\ref{sec:fairness} assesses fairness at both the distribution level (CMI) and in downstream models (EO). Finally, Section~\ref{sec:stats} reports statistical significance tests to provide inferential evidence beyond descriptive comparisons.

\subsubsection{Maximizing the Proxy MI Objective}
\label{sec:q}

\begin{figure}
    \centering
    \subcaptionbox{\adult\label{fig:q-adult}}[0.5\linewidth]{\includegraphics[width=\linewidth]{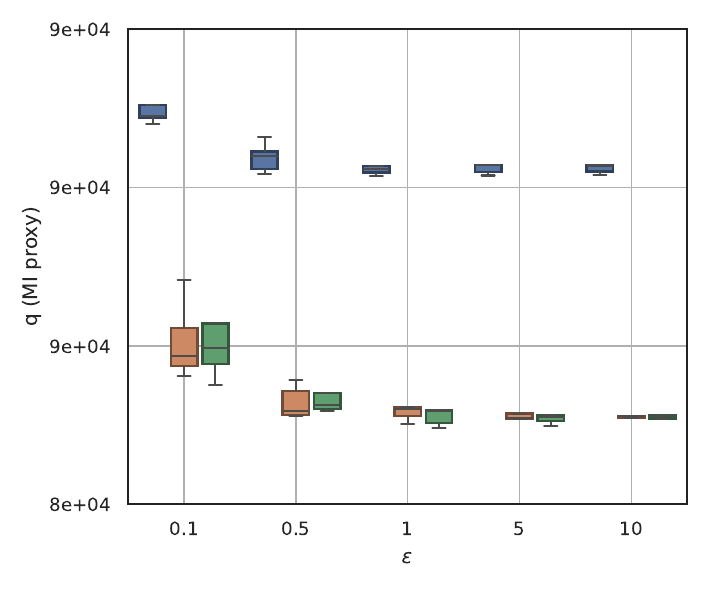}}%
\hfill
    \subcaptionbox{\compas\label{fig:q-compas}}[0.5\linewidth]{\includegraphics[width=\linewidth]{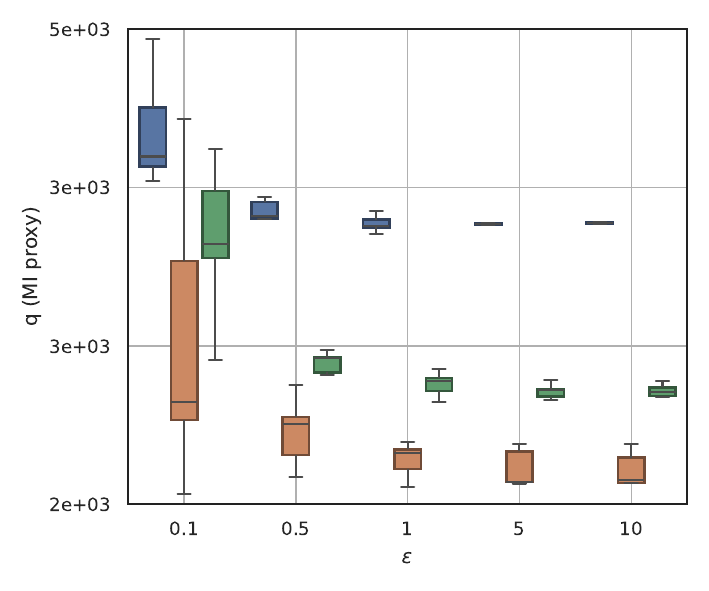}}
    \subcaptionbox{\dutch\label{fig:q-dutch}}[0.5\linewidth]{\includegraphics[width=\linewidth]{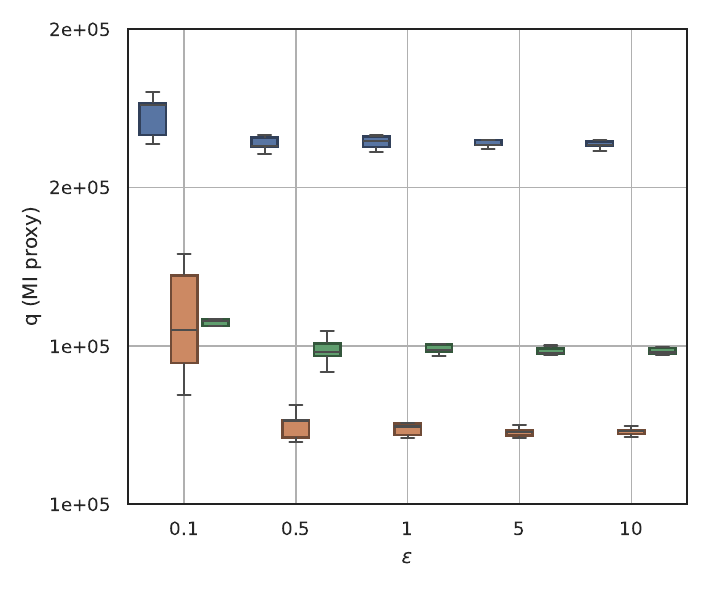}}%
\hfill
    \subcaptionbox{\law\label{fig:q-law}}[0.5\linewidth]{\includegraphics[width=\linewidth]{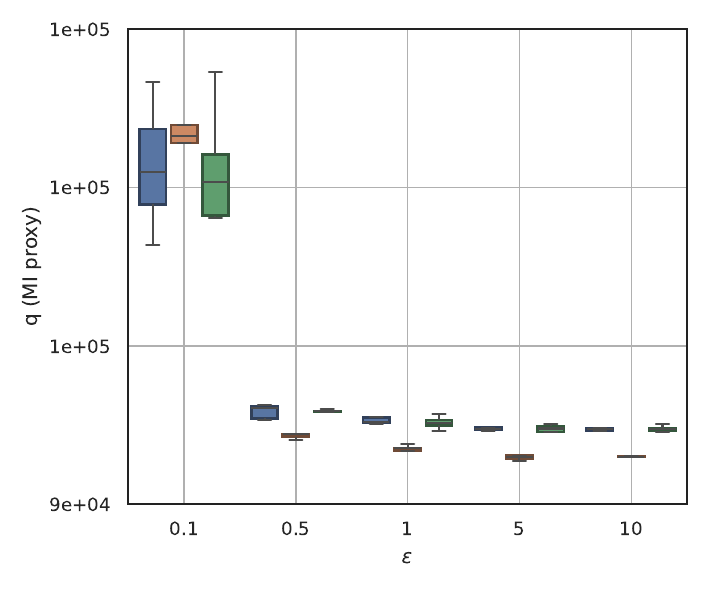}}
    \caption{Total proxy MI score $q$ of selected tree structures.}
    \label{fig:q}
\end{figure}

Figure~\ref{fig:q} reports the total sum of the proxy mutual-information score $q$ over the selected tree edges. 
As expected, \mst achieves the highest objective value, since it applies unconstrained Kruskal’s algorithm and therefore finds the optimal tree under $q$.

Comparing \privci and \prefair, we observe that \privci selects trees with higher total $q$ on COMPAS, Dutch, and Law. 
On Adult, the two methods perform similarly. 
This behavior reflects the difference in design: \privci performs greedy, score-driven edge selection subject to the CI constraint, whereas \prefair removes predefined edges before tree construction. 
In Adult, the removed edge coincides with the fairness-constrained optimum, leading to similar performance.

Across all datasets, increasing the privacy budget $\epsilon$ improves the objective value. 
This is expected, as larger $\epsilon$ reduces noise in the exponential mechanism, resulting in more accurate edge selection.

\subsubsection{Distributional Fidelity}
\label{sec:fidelity}

\begin{figure}
    \centering
    \subcaptionbox{\adult\label{fig:kl-adult}}[0.5\linewidth]{\includegraphics[width=\linewidth]{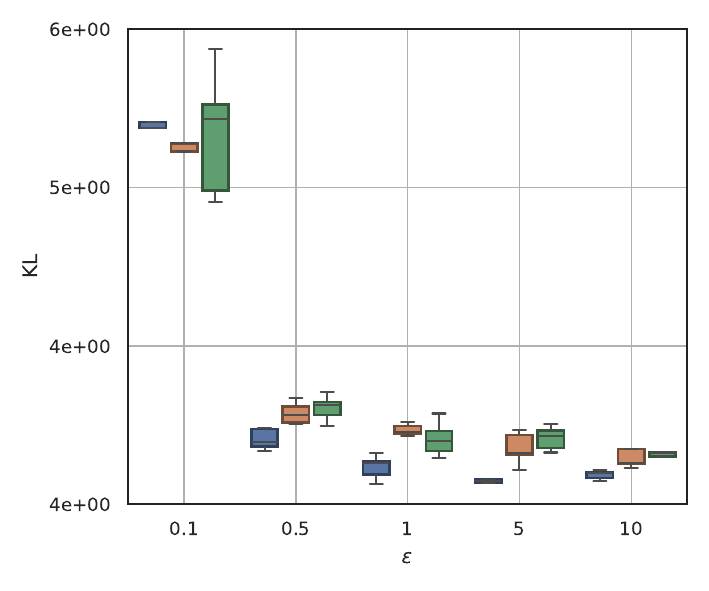}}%
\hfill
    \subcaptionbox{\compas\label{fig:kl-compas}}[0.5\linewidth]{\includegraphics[width=\linewidth]{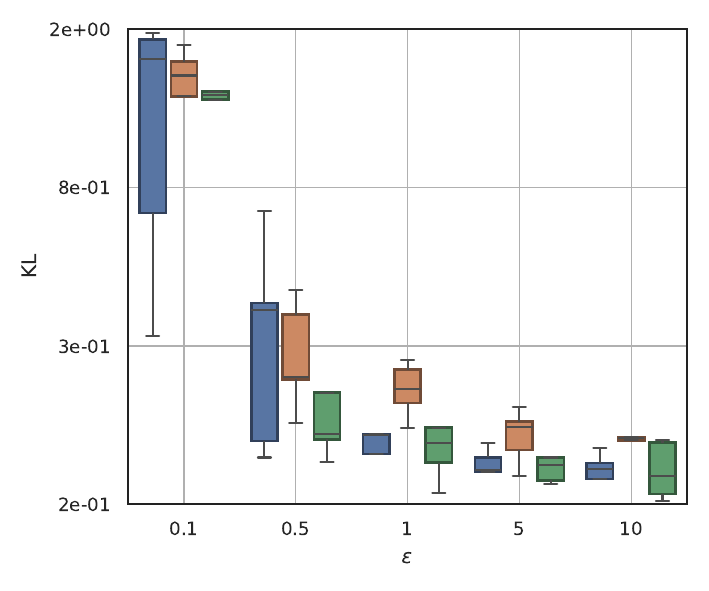}}
    \subcaptionbox{\dutch\label{fig:tv-dutch}}[0.5\linewidth]{\includegraphics[width=\linewidth]{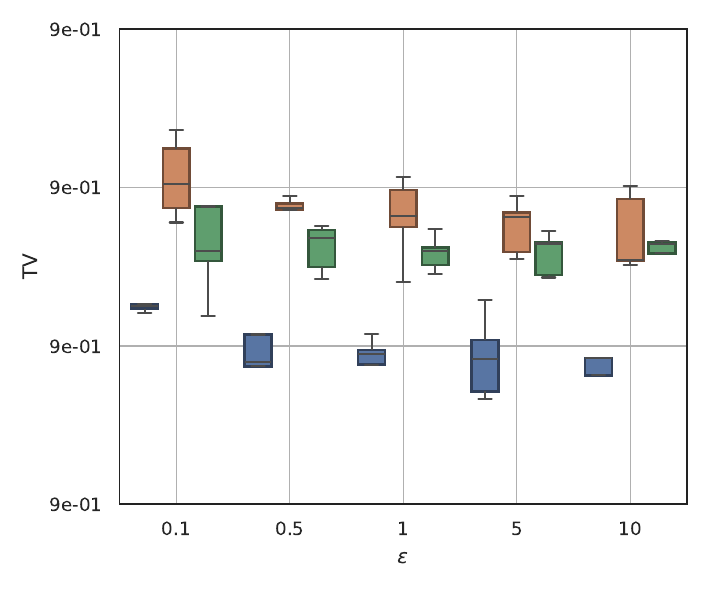}}%
\hfill
    \subcaptionbox{\compas\label{fig:tv-compas}}[0.5\linewidth]{\includegraphics[width=\linewidth]{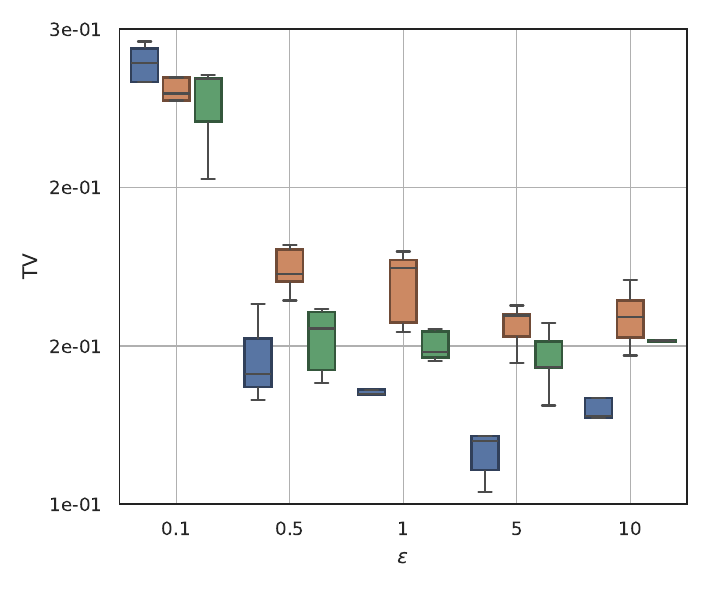}}
    \caption{Distributional fidelity measured by KL divergence and TV distance.}
    \label{fig:fidelity}
\end{figure}

We measure distributional fidelity between the real and synthetic data using KL divergence and total variation (TV) distance. 
Figure~\ref{fig:fidelity} reports these metrics across privacy budgets.

On Adult, \privci and \prefair exhibit similar fidelity, consistent with their similar structural selections. 
On COMPAS and Dutch, \privci produces synthetic distributions closer to the real data. 
As expected, \mst achieves the lowest divergence overall due to the absence of fairness constraints.

Increasing $\epsilon$ improves fidelity for all methods, reflecting reduced noise in private measurements.

\subsubsection{Downstream Predictive Performance}
\label{sec:ml}

\begin{figure}
    \centering
    \subcaptionbox{\adult\label{fig:auc-adult}}[0.5\linewidth]{\includegraphics[width=\linewidth]{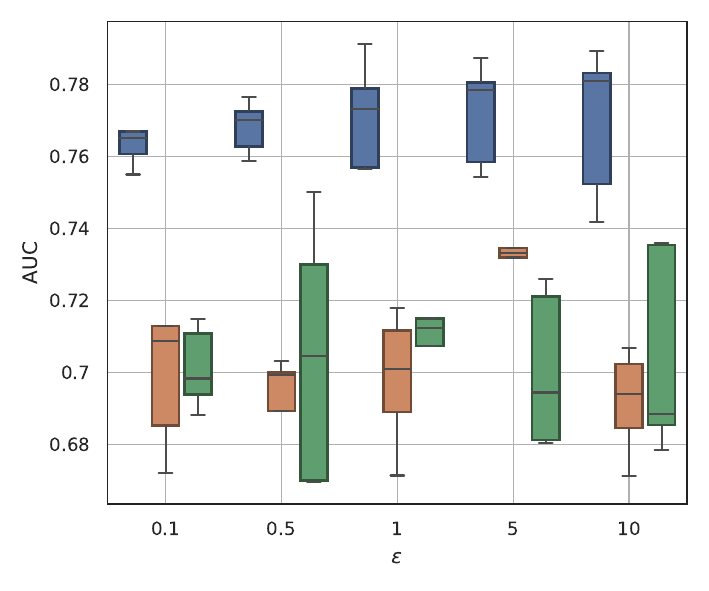}}%
\hfill
    \subcaptionbox{\compas\label{fig:auc-compas}}[0.5\linewidth]{\includegraphics[width=\linewidth]{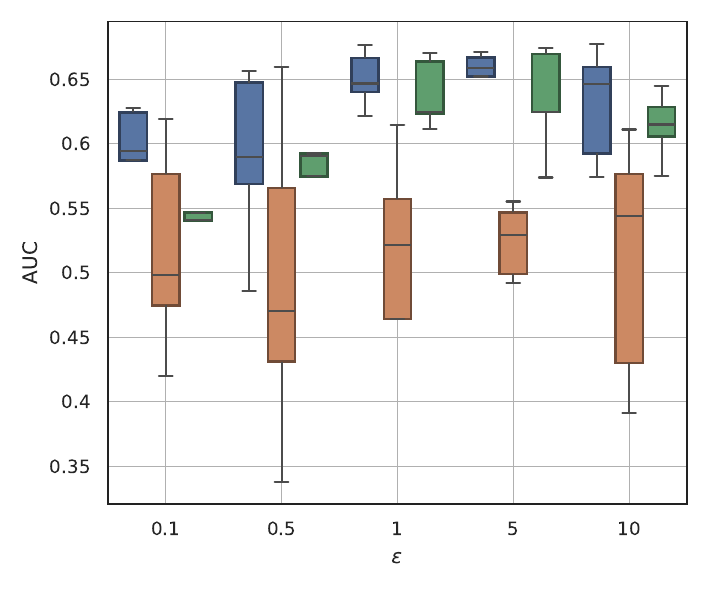}}
    \subcaptionbox{\dutch\label{fig:auc-dutch}}[0.5\linewidth]{\includegraphics[width=\linewidth]{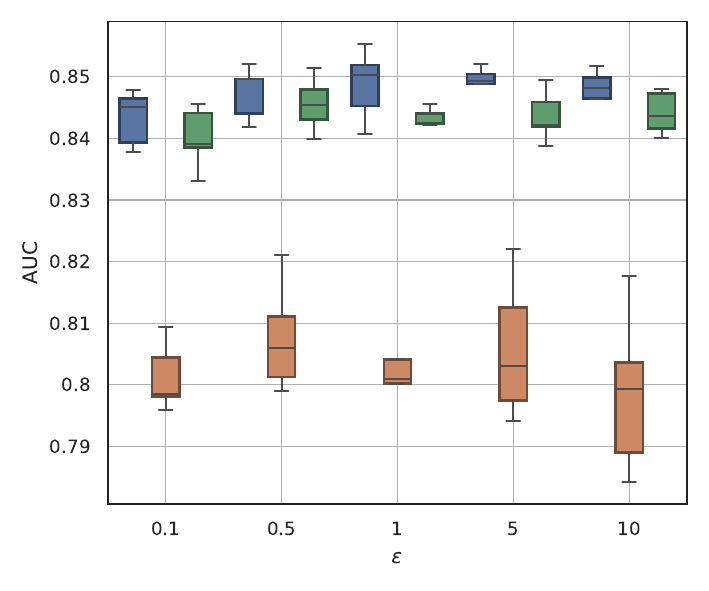}}%
\hfill
    \subcaptionbox{\law\label{fig:auc-law}}[0.5\linewidth]{\includegraphics[width=\linewidth]{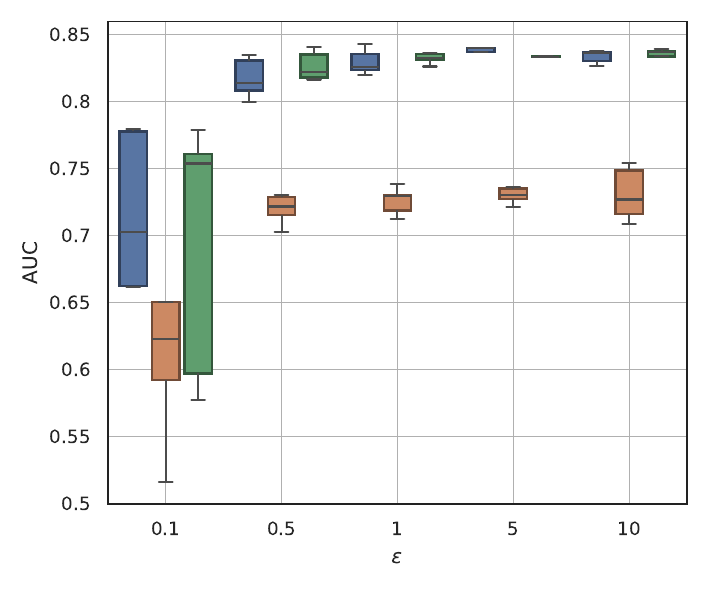}}
    \caption{Downstream predictive performance (AUC).}
    \label{fig:auc}
\end{figure}

To evaluate whether fidelity improvements translate into downstream utility, we train logistic regression models on synthetic data and evaluate AUC on held-out test folds. 
Figure~\ref{fig:auc} reports results across five folds.

With the exception of Adult, \privci consistently achieves higher AUC than \prefair. 
The gap is most pronounced on COMPAS and Law. 
Lower privacy budgets reduce predictive performance for all methods, reflecting the expected privacy–utility tradeoff.

\subsubsection{Fairness at the Distribution and Model Levels}
\label{sec:fairness}

We next evaluate whether utility gains compromise fairness.

\begin{figure}
    \centering
    \subcaptionbox{\adult\label{fig:cmi-adult}}[0.5\linewidth]{\includegraphics[width=\linewidth]{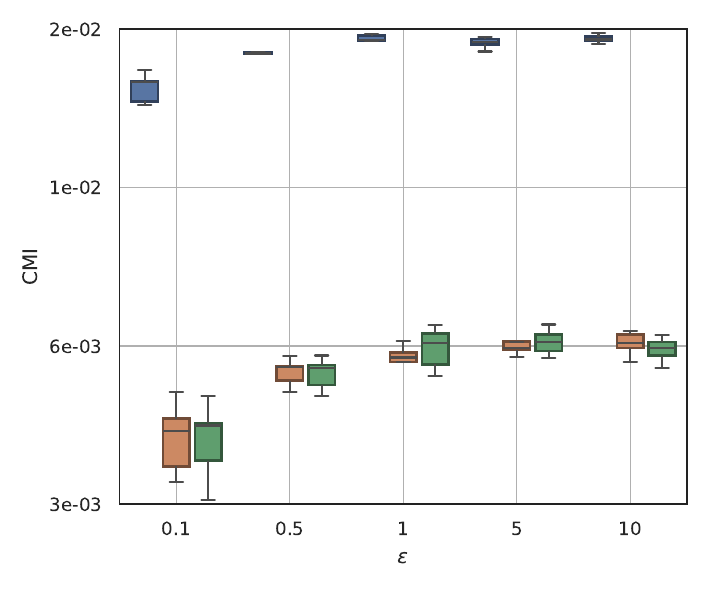}}%
\hfill
    \subcaptionbox{\compas\label{fig:cmi-compas}}[0.5\linewidth]{\includegraphics[width=\linewidth]{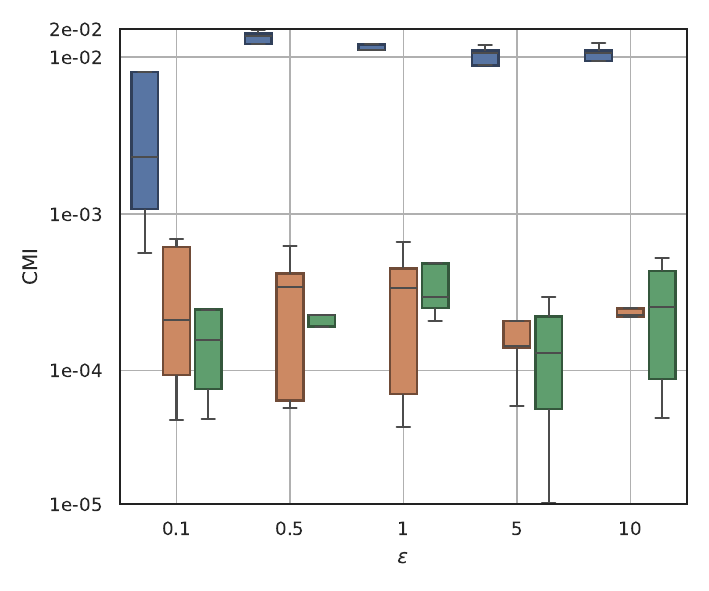}}
    \subcaptionbox{\dutch\label{fig:cmi-dutch}}[0.5\linewidth]{\includegraphics[width=\linewidth]{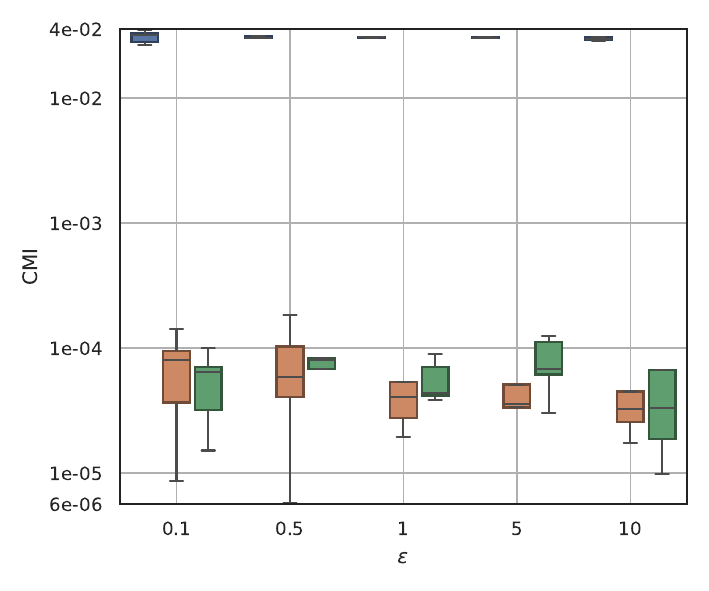}}%
\hfill
    \subcaptionbox{\law\label{fig:cmi-law}}[0.5\linewidth]{\includegraphics[width=\linewidth]{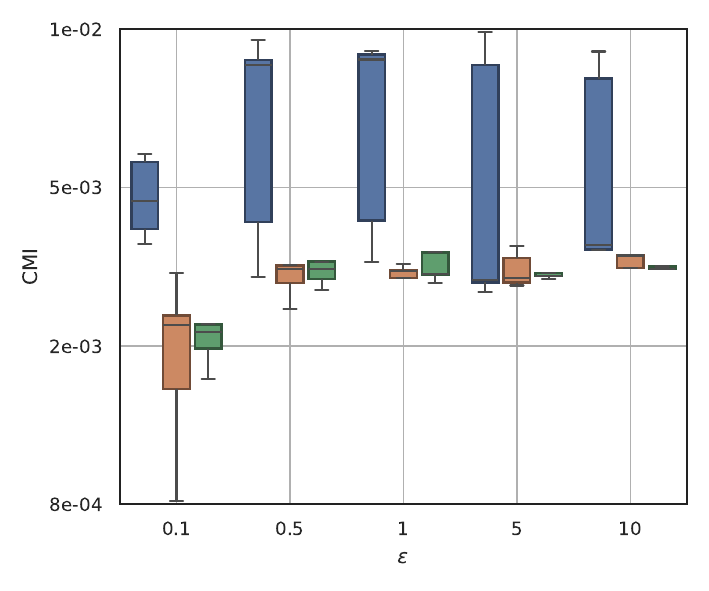}}
    \caption{Conditional mutual information (CMI).}
    \label{fig:cmi}
\end{figure}

Figure~\ref{fig:cmi} reports conditional mutual information (CMI) between protected and outcome variables in synthetic data. 
Both \privci and \prefair substantially reduce CMI relative to \mst, confirming effective enforcement of the CI constraint.

\begin{figure}
    \centering
    \subcaptionbox{\adult\label{fig:eo-adult}}[0.5\linewidth]{\includegraphics[width=\linewidth]{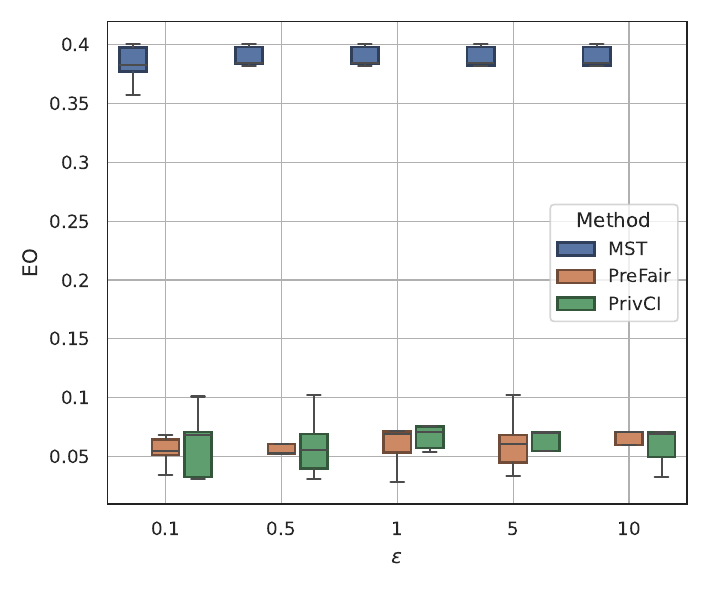}}%
\hfill
    \subcaptionbox{\compas\label{fig:eo-compas}}[0.5\linewidth]{\includegraphics[width=\linewidth]{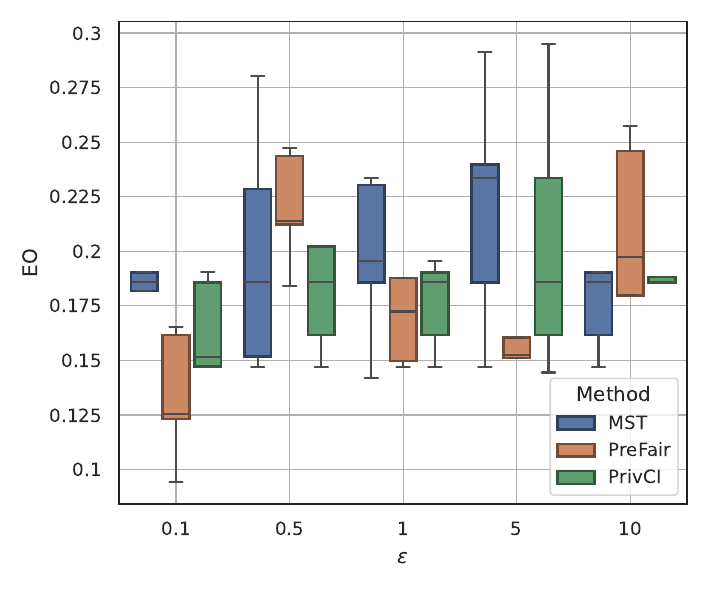}}
    \subcaptionbox{\german\label{fig:eo-german}}[0.5\linewidth]{\includegraphics[width=\linewidth]{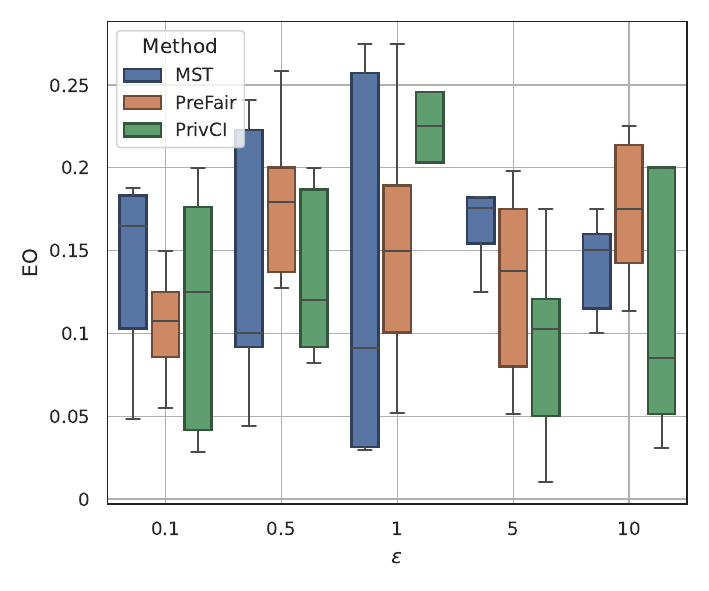}}%
\hfill
    \subcaptionbox{\law\label{fig:eo-law}}[0.5\linewidth]{\includegraphics[width=\linewidth]{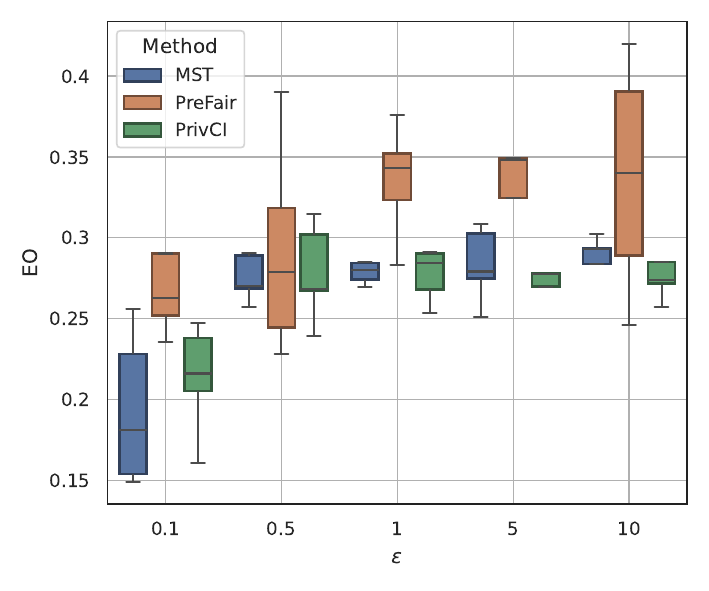}}
    \caption{Equzlied Odds (OD).}
    \label{fig:eo}
\end{figure}

Figure~\ref{fig:eo} shows equalized odds (EO) differences in downstream models. 
Both fairness-aware methods reduce EO disparities compared to \mst. 
However, improvements in EO are dataset-dependent, reflecting the well-known fact that removing distributional dependence does not guarantee improvements in all downstream fairness metrics.

\begin{table*}[t]
\centering
\renewcommand{\arraystretch}{1}
\setlength{\tabcolsep}{2.5mm}
\resizebox{\textwidth}{!}{
\normalsize
\begin{tabular}{lcc cc cc cc cc}
\toprule
\multirow{2}{*}{\textbf{Metric}} 
& \multicolumn{2}{c}{\adult} 
& \multicolumn{2}{c}{\compas}
& \multicolumn{2}{c}{\dutch}
& \multicolumn{2}{c}{\german}
& \multicolumn{2}{c}{\law} \\
\cmidrule{2-11}
 & $\Delta$ & \textbf{$p$-value}
 & $\Delta$ & \textbf{$p$-value}
 & $\Delta$ & \textbf{$p$-value}
 & $\Delta$ & \textbf{$p$-value}
 & $\Delta$ & \textbf{$p$-value} \\
\midrule

q (MI proxy) 
& +34.7983 & 0.6542 
& +331.1423 & \checkmark\,0.0000 
& +2804.1045 & \checkmark\,0.0000 
& +1167.3563 & \checkmark\,0.0040 
& +569.4408 & \checkmark\,0.0258 \\

KL  
& +0.0164 & 0.5837 
& -0.0464 & \checkmark\,0.0019 
& -0.0213 & \checkmark\,0.0015 
& -0.0657 & \checkmark\,0.0426 
& -0.0824 & \checkmark\,0.0000 \\

TV 
& -0.0007 & 0.2712 
& -0.0108 & \checkmark\,0.0002 
& -0.0009 & \checkmark\,0.0006 
& -0.0027 & 0.0502 
& -0.0039 & \checkmark\,0.0000 \\

AUC 
& -0.0000 & 0.4895 
& +0.0885 & \checkmark\,0.0001 
& +0.0408 & \checkmark\,0.0000 
& +0.0497 & \checkmark\,0.0123 
& +0.1016 & \checkmark\,0.0000 \\

CMI 
& +0.0000 & 0.5000 
& -0.0000 & 0.2209 
& +0.0000 & 0.8373 
& +0.0007 & 0.6444 
& +0.0002 & 0.7629 \\

EO  
& +0.0029 & 0.7288 
& +0.0125 & 0.8343 
& -0.0218 & \checkmark\,0.0000 
& -0.0130 & 0.2838 
& -0.0531 & \checkmark\,0.0001 \\

\bottomrule
\end{tabular}
}
\caption{Per-dataset one-sided paired Wilcoxon tests comparing \textbf{PrivCI} against \textbf{PreFair} using fold-level paired units (typically $n=25$). The reported $\Delta$ is the average paired difference $\Delta=\text{PrivCI}-\text{PreFair}$ (negative favors PrivCI for KL/TV/CMI/EO; positive favors PrivCI for AUC and q). A \checkmark\ indicates statistical significance ($p<0.05$) under the appropriate one-sided alternative.}
\label{tab:privci-vs-prefair-all-datasets}
\end{table*}

\subsubsection{Statistical Significance Analysis}
\label{sec:stats}

To quantify whether \privci outperforms \prefair, we use paired one-sided Wilcoxon signed-rank tests. For each dataset, fold, and privacy level $\epsilon$, we compute
\[
d_i = m_i(\text{PrivCI}) - m_i(\text{PreFair}),
\]
and report the average paired effect
\[
\Delta = \frac{1}{n}\sum_{i=1}^n d_i.
\]

Directionality is metric-specific. For KL, TV, CMI, and EO (lower is better), we test $H_1:d_i<0$, so negative $\Delta$ favors \privci. For AUC and $q$ (higher is better), we test $H_1:d_i>0$, so positive $\Delta$ favors \privci. We report one-sided $p$-values and declare significance at $p<0.05$. \privci shows significant gains in fidelity (KL, TV) and predictive performance (AUC) on most datasets. EO improves significantly on selected datasets, whereas CMI differences are not statistically significant. On Adult, effects are generally small, consistent with the descriptive analysis.

\subsection{Discussion and Takeaways} \label{sec:discussion}

Our results support three main conclusions.

First, enforcing CI constraints during measurement can improve utility without weakening fairness enforcement. 
Across datasets and privacy budgets, \privci consistently matches \prefair on CI satisfaction (CMI is statistically indistinguishable), while achieving higher proxy MI $q$, better distributional fidelity (KL/TVD), and higher downstream AUC on most datasets. 
This indicates that structural feasibility filtering during tree construction can preserve more informative dependencies than pre-specifying edge removals.

Second, the strongest gains appear in datasets where the fairness constraint admits multiple feasible trees with materially different utility. 
On COMPAS, Dutch, German, and Law, \privci more often selects higher-scoring CI-feasible structures, which translates into improved fidelity and predictive performance. 
On Adult, effects are small and largely non-significant. This suggests that when the CI-constrained optimum is close to the structure obtained by \prefair, there is limited headroom for improvement.

Third, distribution-level CI enforcement reduces dependence but does not guarantee uniform improvements in downstream fairness metrics. 
Both fairness-aware methods substantially reduce CMI compared to unconstrained \mst, confirming effective structural enforcement of $O \perp S \mid A$. 
EO improvements, however, are dataset-dependent: \privci yields significant EO gains on Dutch and Law but not consistently elsewhere. 
This aligns with prior observations that different fairness metrics capture different notions of disparity, and removing conditional dependence in the data does not necessarily optimize a specific classifier-level metric.

These results indicate that \privci achieves a favorable privacy--utility trade-off among CI-constrained DP synthesizers, matching the privacy accounting of \mst while structurally enforcing the CI constraint and improving utility over the closest baseline on most datasets and privacy budgets.
\section{Conclusion and Future Work}
\label{sec:conclusion}

We presented \priCI, a differentially private synthetic data generation method that enforces conditional independence (CI) constraints during structure selection. By integrating CI feasibility checks into Kruskal-style MST construction, \priCI preserves the privacy accounting and scalability of the standard MST + PrivatePGM pipeline while embedding fairness constraints directly in measurement.

Across benchmark datasets and privacy budgets, \priCI improves structural objective values, distributional fidelity, and downstream predictive performance relative to \prefair, while enforcing the intended CI constraint at the distribution level. These results show that structural separation during measurement can improve the privacy–utility trade-off without weakening fairness guarantees.

Several extensions remain open, including handling multiple or overlapping CI constraints, scaling beyond tree-structured models, and supporting mixed-type or continuous domains, as well as extending CI-aware enforcement to DP-trained deep generative models. More broadly, an important direction is to move beyond hard feasibility constraints and jointly optimize privacy and fairness within the mechanism itself—for example, through fairness-aware budget allocation and integrating CI as soft penalties in the selection objective. This would enable a principled trade-off between privacy, utility, and fairness. Additionally, incorporating group-level or distributional privacy notions and designing downstream-aware synthesis that anticipates model behavior could better align distribution-level CI guarantees with model-level fairness outcomes.

\bibliographystyle{IEEEtran}
\bibliography{bibliography}

\end{document}